\documentclass[jkps,preprint,fleqn,showpacs,showkeys]{revtex4}
\usepackage{graphicx}
\usepackage{amssymb}
\usepackage{amsmath}
\usepackage{epstopdf}
\usepackage{bm}
\begin{document}
\setcounter{page}{1}
\title{Numerical studies on the synchronization of a network of mutually coupled simple chaotic systems}
\author{G. Sivaganesh}
\affiliation{Department of Physics, Alagappa Chettiar Government College of Engineering $\&$ Technology, Karaikudi, Tamilnadu-630 004, India}
\author{A. Arulgnanam}
\email{gospelin@gmail.com}
\affiliation{Department of Physics, St. John's College, Palayamkottai, Tamilnadu-627 002, India}
\author{A. N. Seethalakshmi}
\affiliation{Department of Physics, The M.D.T Hindu College, Tirunelveli, Tamilnadu-627 010, India}


\begin{abstract}
We present in this paper, the synchronization dynamics observed in a network of mutually coupled simple chaotic systems. The network consisting of chaotic systems arranged in a square matrix network is studied for their different types of synchronization behavior. The chaotic attractors of the simple $2 \times 2$ matrix network exhibiting strange non-chaotic attractors in their synchronization dynamics for smaller values of the coupling strength is reported. Further, the existence of islands of unsynchronized and synchronized states of strange non-chaotic attractors for smaller values of coupling strength is observed. The process of complete synchronization observed in the network with all the systems exhibiting strange non-chaotic behavior is reported. The variation of the slope of the singular continuous spectra as a function of the coupling strength confirming the strange non-chaotic state of each of the system in the network is presented. The stability of complete synchronization observed in the network is studied using the {\emph{Master Stability Function}}.
\end{abstract}

\pacs{05.45.Xt, 05.45.-a}

\keywords{Synchronization, Strange non-chaotic attractors, Master stability function}

\maketitle

\section{Introduction}
\label{sec:1}
The observation of a chaotic attractor in the {\emph{Chua's circuit}} \cite{Matsumoto1984} and the implementation of the {\emph{Chua's diode}} using Operational Amplifiers \cite{Kennedy1992} have led to the design of a large number of chaotic circuits. Several simple electronic circuit systems have been identified to exhibit chaotic phenomena in their dynamics  \cite{Inaba1991,Murali1994,Arulgnanam2009,Arulgnanam2015}. An intermediate dynamical state between periodic and chaotic motion without any sensitive dependence on initial conditions and with a fractal nature namely, the Strange Non-chaotic Attractor (SNA), has been observed by Grebogi {\emph{et al} \cite{Grebogi1984}. Several simple chaotic systems have also been found to exhibit SNA behavior upon quasiperiodic forcing \cite{Venkatesan1999,Thamilmaran2006,Arulgnanam2015a,Arulgnanam2015b}. The application of a SNA observed in a a second-order chaotic system, for computation, has been reported recently \cite{Sathish2018}. The second-order, non-autonomous, dissipative chaotic circuit namely, the {\emph{Murali-Lakshmanan-Chua}} (MLC) circuit \cite{Murali1994}, is an important circuit system that possesses several chaotic behaviors in its dynamics. This circuit has been widely studied experimentally, numerically and analytically because of the mathematical simplicity of the circuit equations characterizing the system. Further, the circuit exhibits SNA behavior under the application of quasiperiodic forcing \cite{Venkatesan1999}. Synchronization is the dynamical process observed in coupled chaotic systems \cite{Pecora1990}. The stability of complete synchronization observed in coupled chaotic systems is studied using the {\emph{Master Stability Function}} (MSF)\cite{Pecora1998}. The synchronization behavior observed in unidirectionally and mutually coupled MLC circuits have been studied numerically and analytically \cite{Sivaganesh2017,Sivaganesh2018,Sivaganesh2018a}.\\

We present in this paper, the synchronization dynamics observed in a matrix network of mutually coupled MLC circuits. The mechanism leading to complete synchronization in the $2 \times 2$ matrix network is investigated and reported. The evolution of SNAs in the dynamical process of synchronization of the network is observed. The stability of the stable synchronized states observed in the network is studied using the MSF.

\section{Network of {\emph{Murali-Lakshmanan-Chua}} Circuits}
\label{sec:2}

The {\emph{Murali-Lakshmanan-Chua}} circuit is the simplest second-order non-autonomous electronic circuit system in which chaos is realized. It is a sinusoidally forced series LCR circuit with a nonlinear element namely, the {\emph{Chua's diode}} as shown in figure \ref{fig:1}. The mathematical simplicity of the circuit equations enabled researchers to study the bifurcations and chaos observed in the circuit both numerically and analytically. The prominent chaotic attractors observed in the circuit at the values of the amplitude of external force $f=0.1$ and $f=0.14$ are as shown in  figures \ref{fig:2}(a) and \ref{fig:2}(b), respectively.


\begin{figure}[h]
\centering
\includegraphics[width=0.5\textwidth]{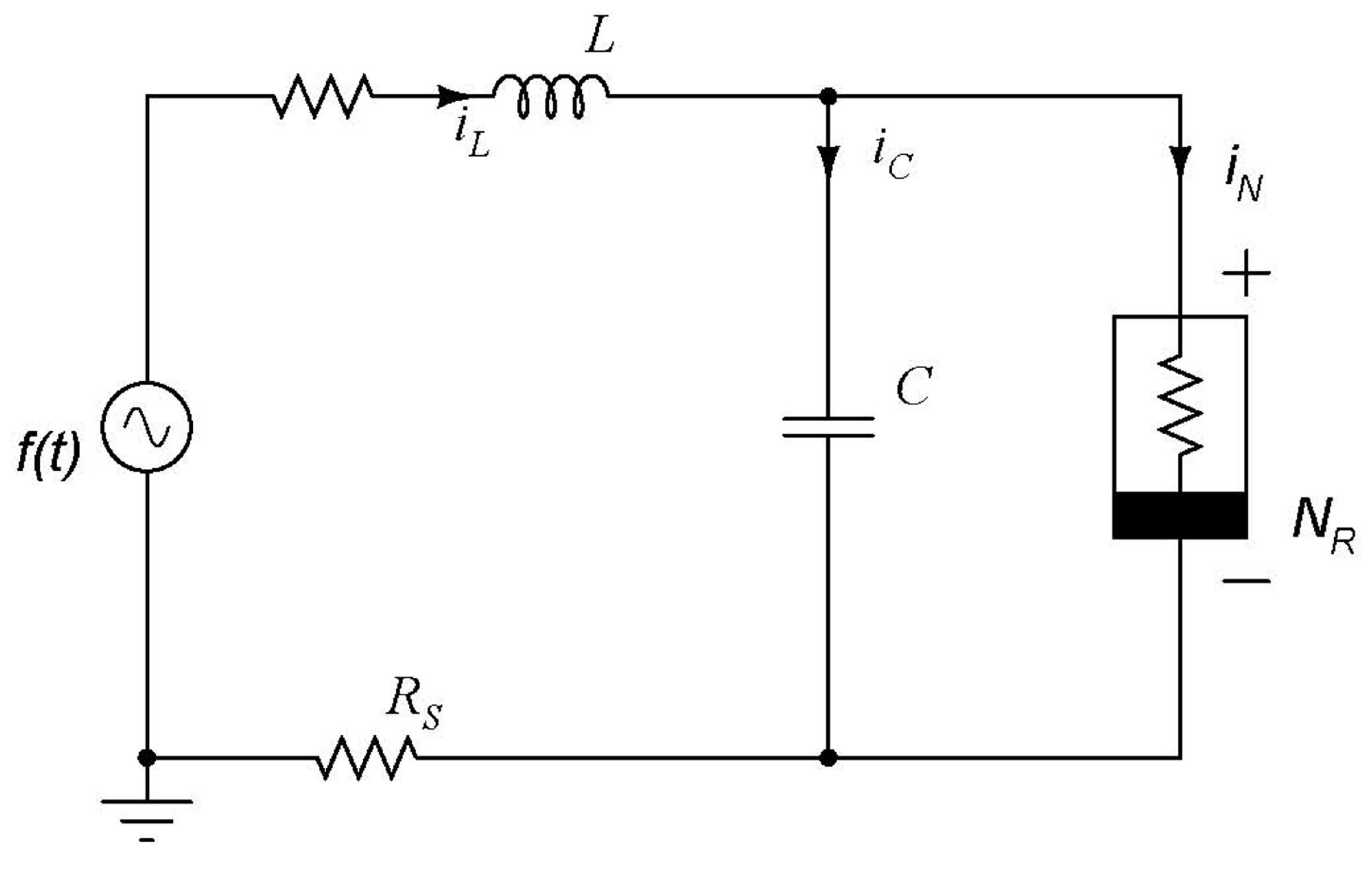}
\caption{(Color Online) Schematic circuit realization of the {\emph{Murali-Lakshmanan-Chua}} circuit. }

\label{fig:1}
\end{figure}


\begin{figure}[h]
\centering
\includegraphics[width=1\textwidth]{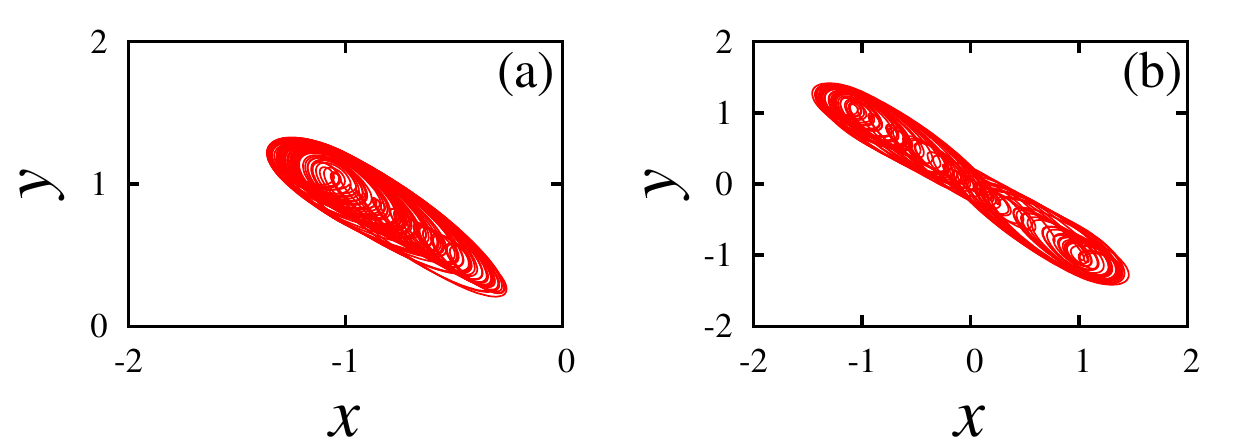}
\caption{(Color Online) (a) One-band and (b) Double-band chaotic attractors of the MLC circuit obtained at the amplitude of the external force $f=0.1$ and $f=0.14$, respectively. }

\label{fig:2}
\end{figure}


\begin{figure}[h]
\centering
\includegraphics[width=0.5\textwidth]{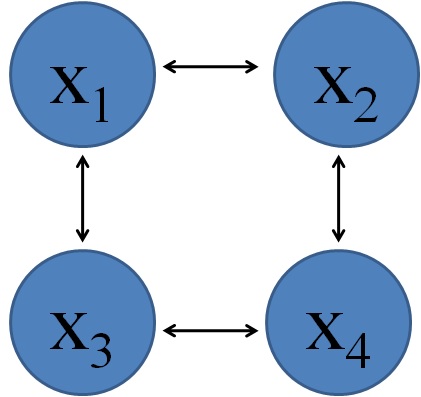}
\caption{(Color Online) Schematic representation of the $2 \times 2$ matrix network. The variables $x_1, x_2, x_3, x_4$ represents the x-variables of each chaotic system mutually-coupled to each other. }

\label{fig:3}
\end{figure}


\begin{figure}[h]
\centering
\includegraphics[width=1\textwidth]{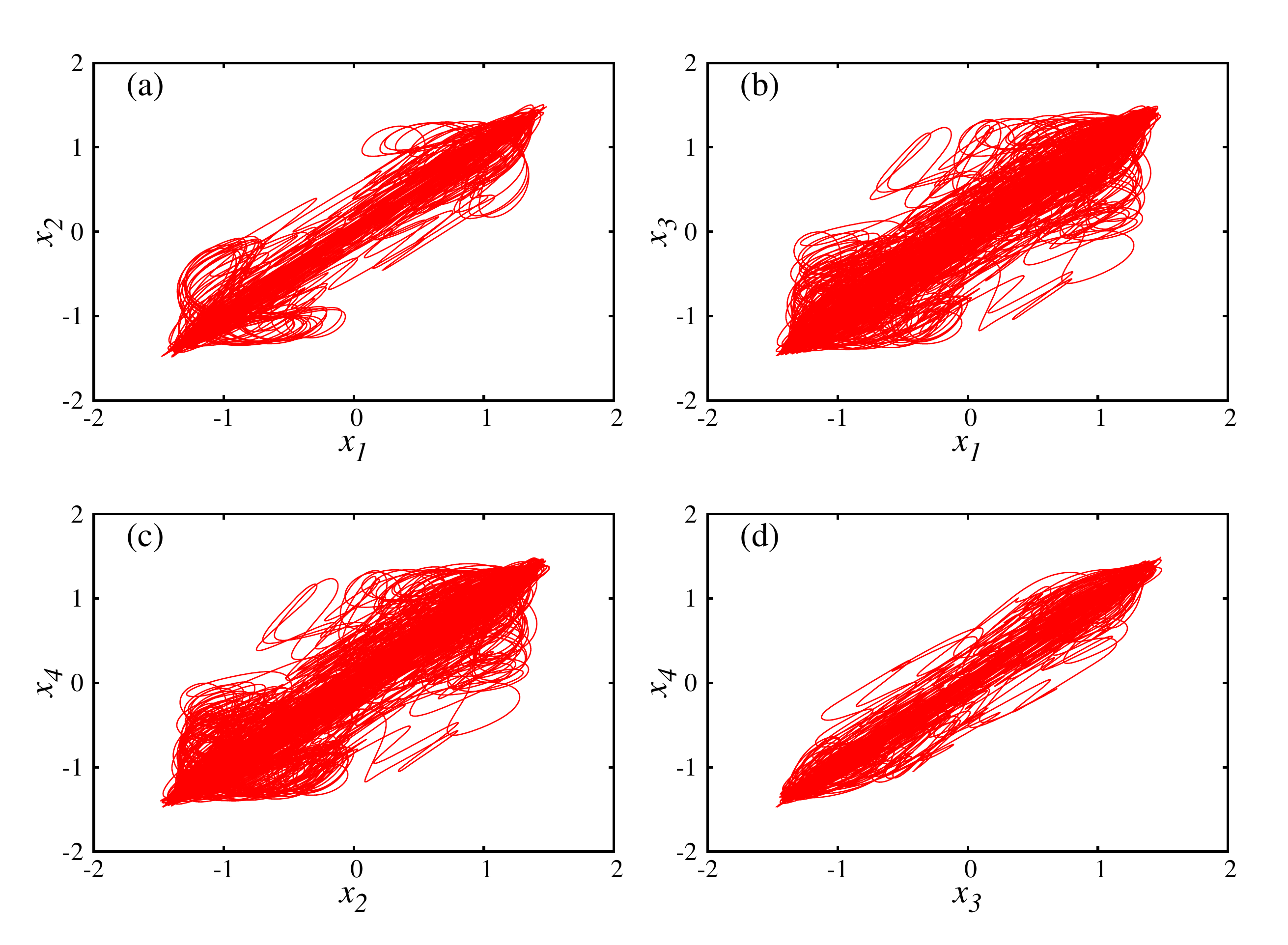}
\caption{(Color online) Phase-portraits of the unsynchronized states of the coupled chaotic systems for the value of the coupling parameter $\epsilon = 0.0188$. All the systems exist in SNA states and are unsynchronized with others.}
\label{fig:4}
\end{figure}


\begin{figure}[h]
\centering
\includegraphics[width=1\textwidth]{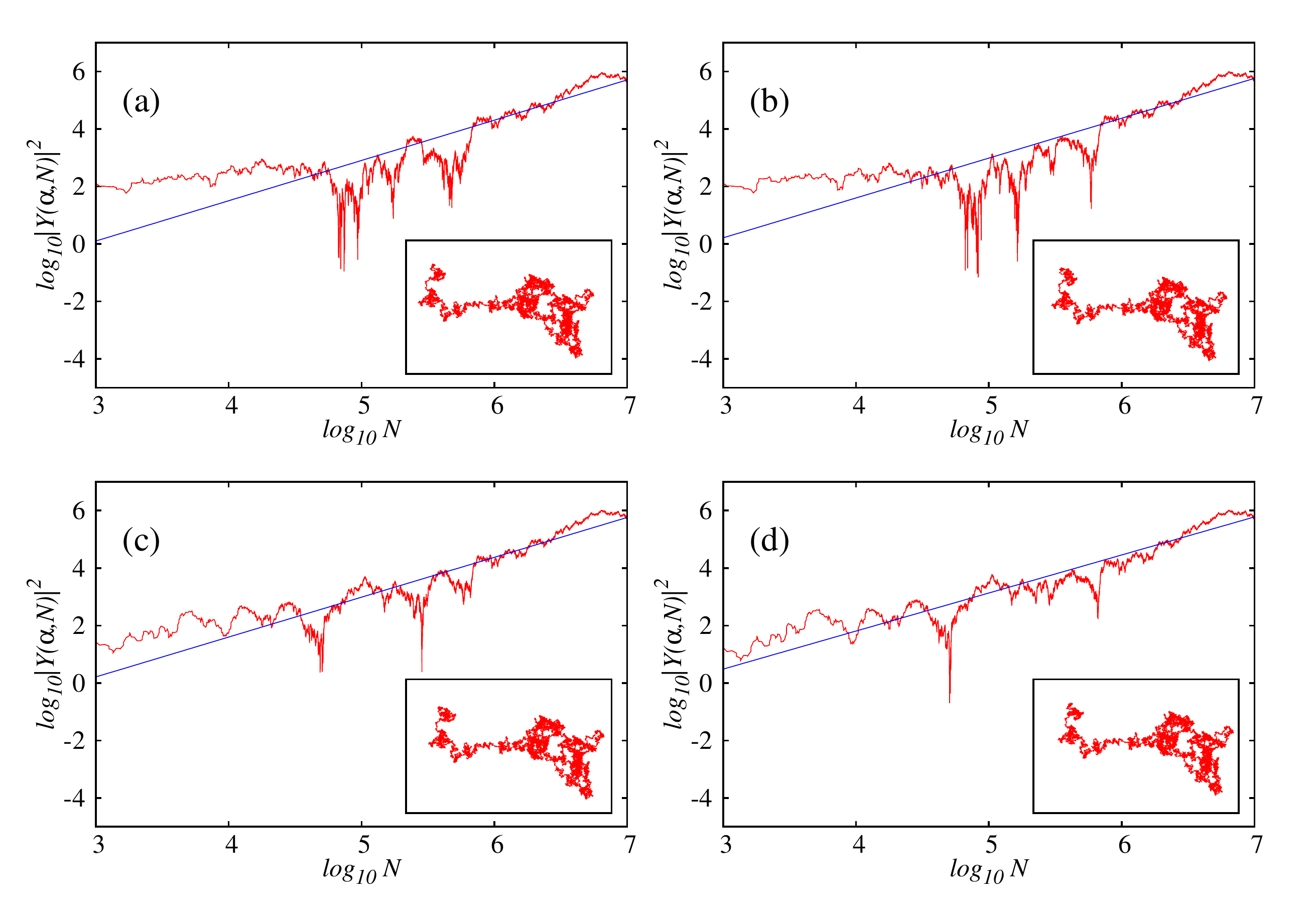}
\caption{(Color Online) (a) Singular continuous spectrum analysis of each system for the control parameter value $\epsilon = 0.0188$ of the systems. Plot of $|Y(\alpha,N)|^2$ vs. $N$ shows the power-law scaling with the inner figure showing the orbit walker in the complex plane exhibiting a fractal nature, suggesting the presence of SNA. The slope of the fractal dimension (a) $\mu_1 = 1.152$ for system 1 for the variable $x_1$, (b) $\mu_2 = 1.0615$ for system 2 for the variable $x_2$, (c) $\mu_3 = 1.1154$ for system 3 for the variable $x_3$, and (d) $\mu_4 = 1.1582$ for system 4 for the variable $x_4$.}

\label{fig:5}
\end{figure}


\begin{figure}[h]
\centering
\includegraphics[width=1\textwidth]{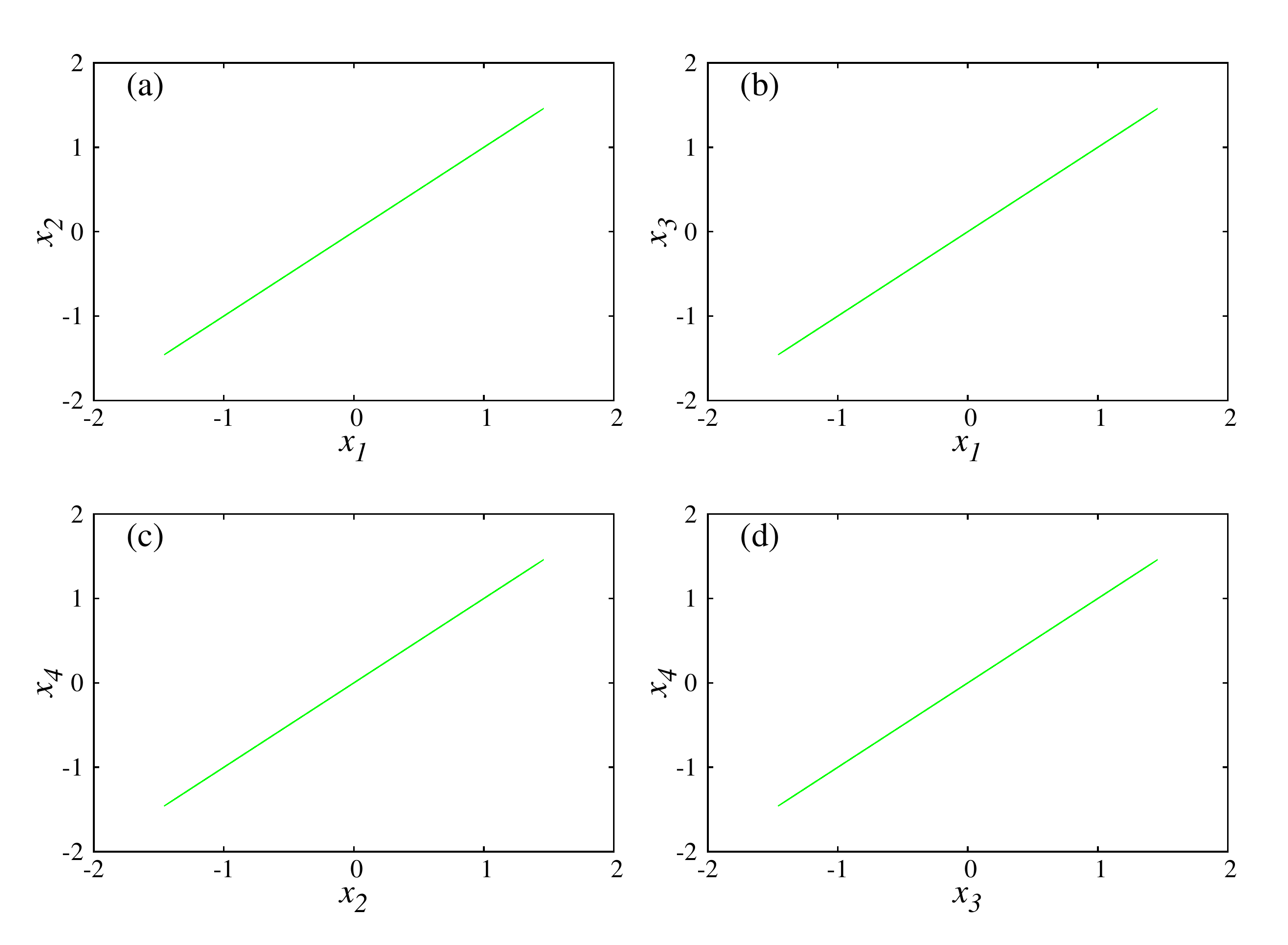}
\caption{(Color online) Phase-portraits of the synchronized states of the coupled chaotic systems for the value of the coupling parameter $\epsilon = 0.02$. All the coupled chaotic systems gets completely synchronized.and all of them exist in the SNA state.}
\label{fig:6}
\end{figure}


\begin{figure}[h]
\centering
\includegraphics[width=1\textwidth]{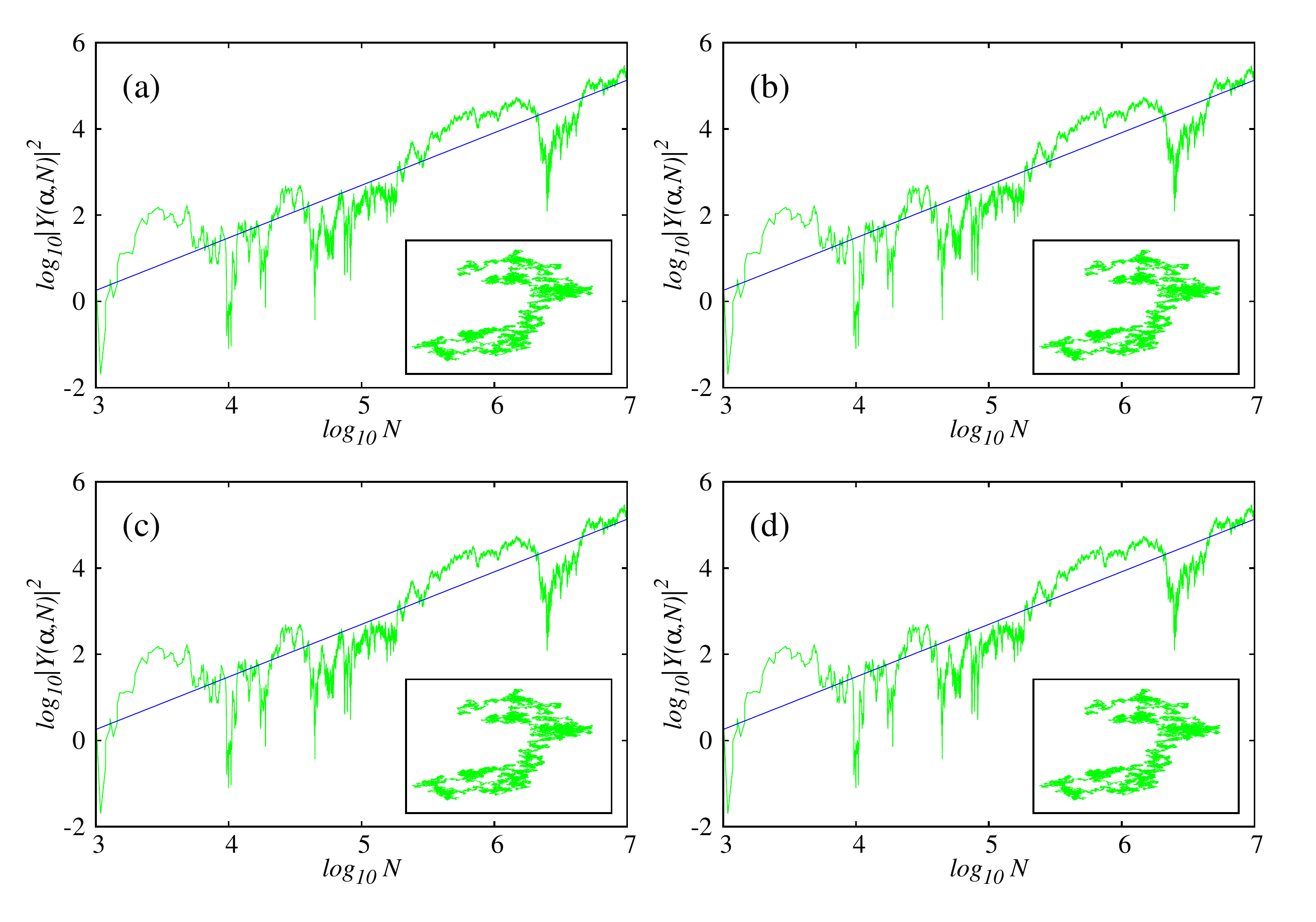}
\caption{(Color Online) (a) Singular continuous spectrum analysis of each system for the control parameter value $\epsilon = 0.02$ of the systems. Plot of $|Y(\alpha,N)|^2$ vs. $N$ shows the power-law scaling with the inner figure showing the orbit walker in the complex plane exhibiting a fractal nature, suggesting the presence of SNA. (a) $\mu_1 = 1.3761$ for system 1 for the variable $x_1$, (b) $\mu_2 = 1.3761$ for system 2 for the variable $x_2$, (c) $\mu_3 = 1.3761$ for system 3 for the variable $x_3$, and (d) $\mu_4 = 1.3761$ for system 4 for the variable $x_4$.}

\label{fig:7}
\end{figure}


\begin{figure}[h]
\centering
\includegraphics[width=0.66\textwidth]{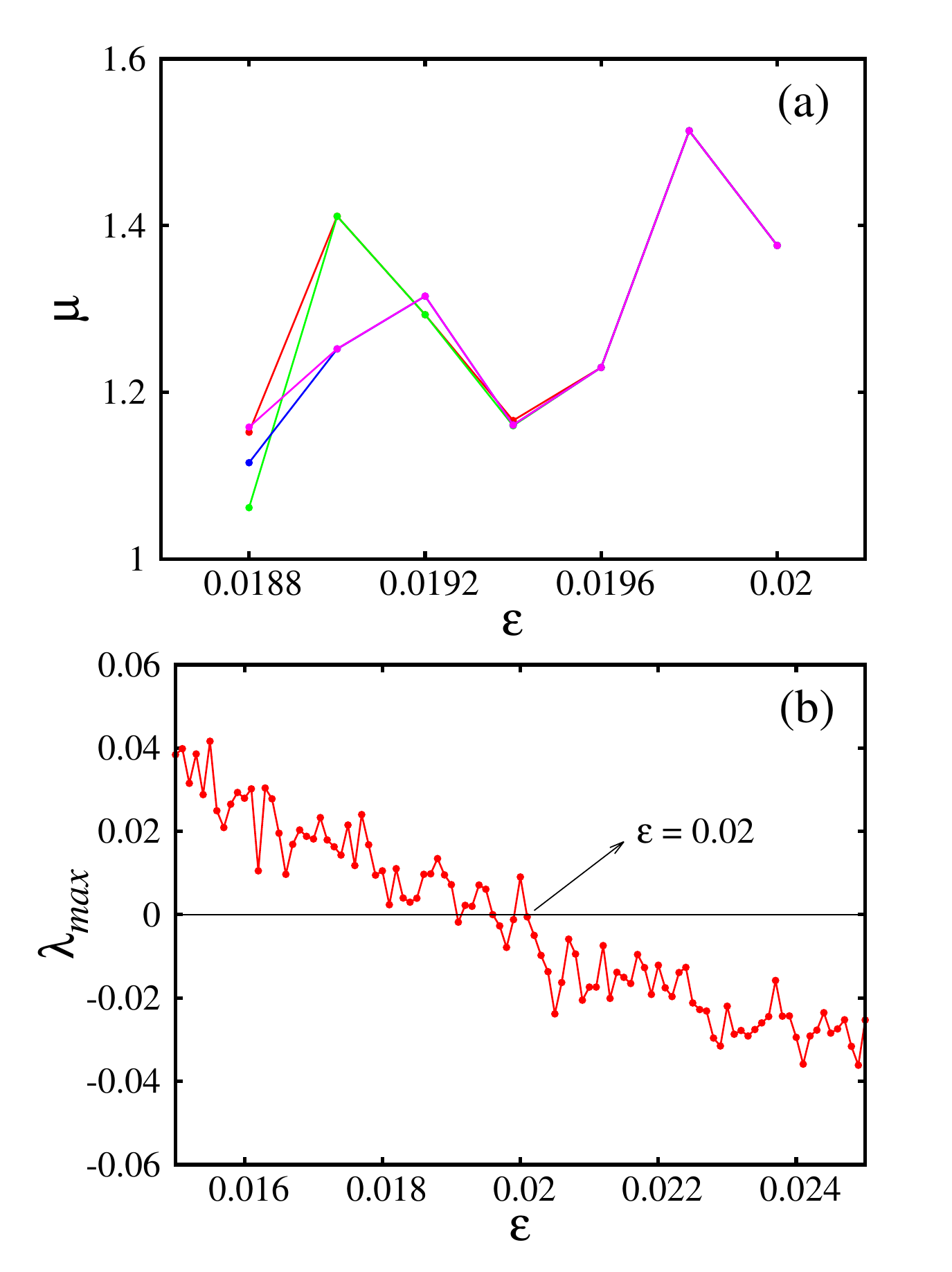}
\caption{(Color Online) (a) Slope ($\mu$) of the $|Y(\alpha,N)|^2$ vs. $N$ plot as function of the coupling parameter for the x-variables of system 1 (red), system 2 (green), system 3 (blue) and system 4 (magenta) indicating the persistence of the systems in the SNA region. (b) MSF of the matrix network indicating the transition of the matrix network to stable synchronized states for $\epsilon \ge 0.02$.}
\label{fig:8}
\end{figure}


\begin{figure}[h]
\centering
\includegraphics[width=0.66\textwidth]{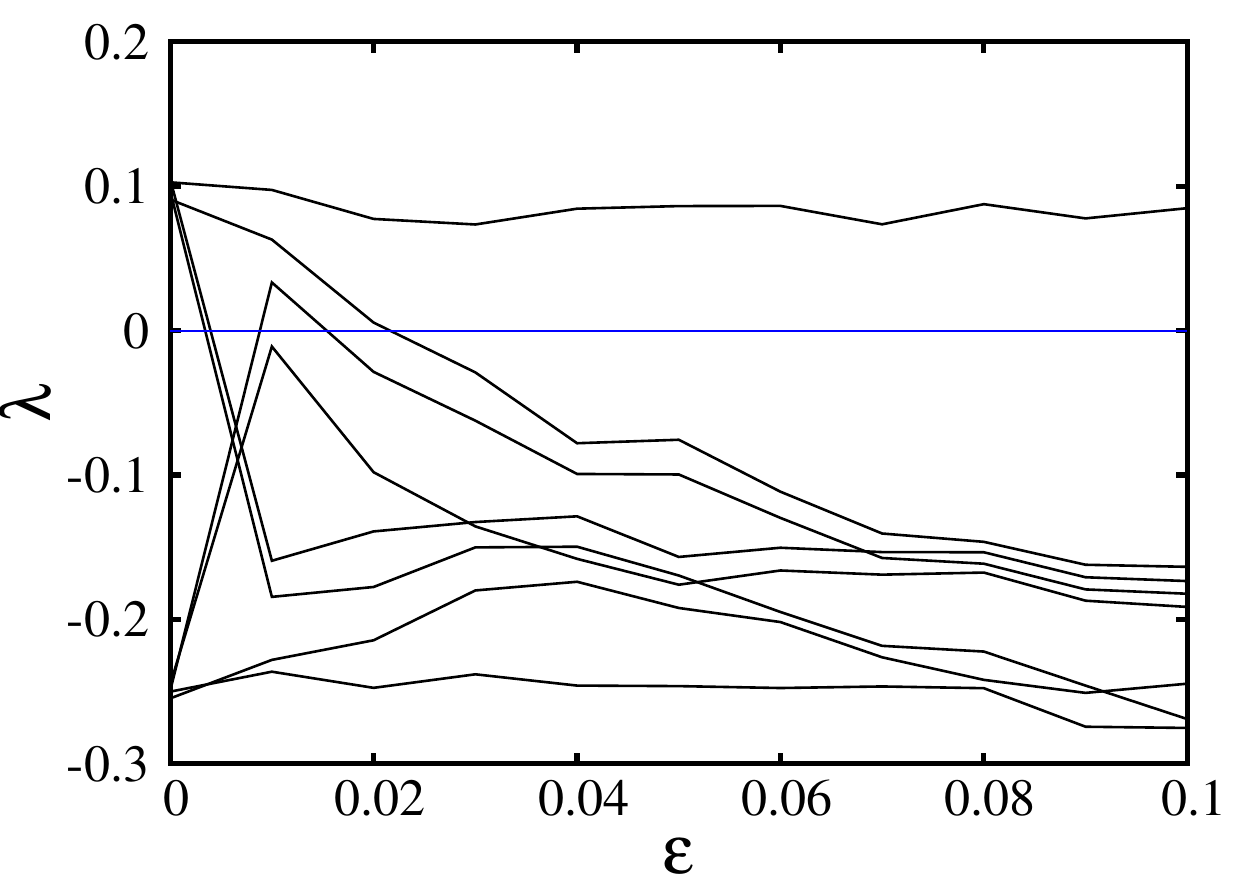}
\caption{The eight largest Lyapunov exponents $(\lambda)$ of the matrix network as a function of the coupling strength $(\epsilon)$.}
\label{fig:9}
\end{figure}


The synchronization behavior of the MLC circuit in a $2 \times 2$ matrix network is presented in this section. The network consists of four MLC circuit systems each system being mutually coupled to its nearest neighbors and hence the systems does not interact diagonally. The circuits are coupled to each other through the x-variable of the system. The schematic diagram of the matrix network of chaotic systems coupled through the x-variables is as shown in figure \ref{fig:3}. Each of the circuit system present in the matrix network is operated at the double-band chaotic state shown in figure \ref{fig:2}(b).

The normalized state equations of each of the chaotic system in the matrix network can be written as

{\bf System 1:}
\begin{eqnarray}
\dot x_1  &=&  y_1 - g(x_1) + \epsilon(x_2-x_1) + \epsilon(x_3-x_1), \\ 
\dot y_1  &=&  -\sigma y_1 - \beta x_1 + f sin(z_1) ,\\ 
\dot z_1  &=&  \omega,
\end{eqnarray}

{\bf System 2:}
\begin{eqnarray}
\dot x_2  &=&  y_2 - g(x_2) + \epsilon(x_1 - x_2) + \epsilon(x_4 - x_2), \\ 
\dot y_2  &=&  -\sigma y_2 - \beta x_2 + f sin(z_2) ,\\ 
\dot z_2  &=&  \omega,
\end{eqnarray}

{\bf System 3:}
\begin{eqnarray}
\dot x_3  &=&  y_3 - g(x_3)+ \epsilon(x_1 - x_3) + \epsilon(x_4 - x_3), \\ 
\dot y_3  &=&  -\sigma y_3 - \beta x_3 + f sin(z_3) ,\\ 
\dot z_3  &=&  \omega,
\end{eqnarray}

{\bf System 4:}
\begin{eqnarray}
\dot x_4  &=&  y_4 - g(x_4)+ \epsilon(x_2 - x_4) + \epsilon(x_3 - x_4), \\ 
\dot y_4  &=&  -\sigma y_4 - \beta x_4 + f sin(z_4) ,\\ 
\dot z_4  &=&  \omega,
\end{eqnarray}

where $g(x_i)$ is the mathematical representation of the piecewise-linear behavior of the {\emph{Chua's diode}} given as
\begin{equation}
g(x_i) =
\begin{cases}
bx_i+(a-b) & \text{if $x_i \ge 1$}\\
ax_i & \text{if $|x_i|\le 1$}\\
bx_i-(a-b) & \text{if $x_i \le -1$}
\end{cases}
\end{equation}

where i=1,2,3,4. The normalized parameters of the circuit take the values $a=-1.02,~b=-0.55,~\beta = 1.0,~\nu = 0.015$ and $\omega = 0.72$.\\

The synchronization dynamics of the network is studied by varying the strength of the coupling parameter. The chaotic systems present at each node of the network operating at different initial conditions are unsynchronized when the coupling parameter $\epsilon = 0$. Hence, the individual systems in the network becomes independent of each other and does not synchronize with its neighboring systems. With the increase in the coupling parameter $\epsilon$, the chaotic systems at each node of the network becomes coupled to each other and approaches the synchronized state. However, the mechanism of synchronization observed in this simple network makes an interesting study of this simple network. The interesting phenomenon being the emergence of strange non-chaotic attractors (SNAs) in the dynamics of the coupled systems en-route complete synchronization. With increase in the coupling parameter $\epsilon$, the chaotic attractors at each node evolves into an SNA. All the  attractors in the network evolve into SNA's at the coupling parameter value $\epsilon = 0.0188$. The SNA at each node is unsynchronized with the attractor in its neighboring node as shown in figure \ref{fig:4}. The SNA behavior of the attractors at each node is confirmed through slopes of the singular continuous spectrum (SCS) and random walk motion in the complex plane as shown in figure \ref{fig:5}. The slopes of the singular continuous spectrum obtained for each system are $\mu_1=1.152,~\mu_2=1.0615,~\mu_3=1.1154$ and $\mu_4=1.1582$. With the increase in the coupling parameter to the value $\epsilon = 0.019$, the chaotic attractors completely synchronizes along the rows and gets phase synchronized along the columns. However, all the systems in the network exists in the SNA state. The SNA behavior of the attractors at each node is confirmed through the values of the slopes of the singular continuous spectrum given as $\mu_{1,2} = 1.4112$ and $\mu_{3,4} = 1.252$. Further increase in the coupling parameter to the value $\epsilon = 0.02$ results in complete synchronization (CS) of all the systems in the matrix network as shown in figure \ref{fig:6}. It has to be noted that all the systems still remain in the SNA state with the slopes of the singular continuous spectra given as $\mu_{1,2,3,4} = 1.3761$. The SNA behavior of the attractors at each node is confirmed through singular continuous spectrum analysis and random walk motion in the complex plane as shown in figure \ref{fig:7}. \\

The variation of the slope $(\mu)$ of the $|Y(\alpha,N)|^2$ vs. $N$ plot of each of the attractor present in the network as a function of the coupling parameter ($\epsilon$) shown in figure \ref{fig:8}(a) indicates the existence of all the systems of the matrix at the SNA state and the synchronization of SNA's in the adjacent nodes. The convergence of the slopes to a single value $(\mu = 1.3761)$ confirms the complete synchronization state of all the SNA's at the value of the coupling parameter $\epsilon=0.02$.
The stability of synchronization of the mutually-coupled systems in the matrix network is observed through the {\emph{Master Stability Function}}. Figure \ref{fig:8}(b) showing the MSF of the coupled network indicates the transition of the MSF to negative values for $\epsilon \ge 0.02$ and hence confirms the state of complete synchronization observed in the network. The largest Lyapunov exponents of the network obtained as a function of the coupling parameter shown in figure \ref{fig:9} indicates the transition of the network to stable synchronized states.

\section{Conclusion}
\label{sec:2}

In this report we have presented the synchronization dynamics of a simple $2 \times 2$ matrix network of chaotic systems. The network consisting of an array of chaotic attractors pertaining to a simple chaotic circuit exhibit islands of unsynchronized and synchronized states in its dynamics. The significant feature has been the evolution of SNA's and their persistence in the entire synchronization dynamical process of the network. The emergence, propagation and synchronization of SNA's observed in a matrix network of simple chaotic systems is reported in the literature for the first time. The present study can be further extended for larger matrix networks of chaotic systems for the identification and generalization of the hidden phenomenon to achieve complete synchronization process through emergence of strange non-chaotic attractors.

\end{document}